\documentclass[a4paper,12pt]{article}
\usepackage{epsfig}
\usepackage{amssymb}
\usepackage{cite}

\newcommand{\be}{\begin{equation}}
\newcommand{\ee}{\end{equation}}
\newcommand{\br}{\begin{eqnarray}}
\newcommand{\er}{\end{eqnarray}}
\newcommand{\ba}{\begin{array}}
\newcommand{\ea}{\end{array}}
\newcommand{\bi}{\begin{itemize}}
\newcommand{\ei}{\end{itemize}}
\newcommand{\bn}{\begin{enumerate}}
\newcommand{\en}{\end{enumerate}}
\newcommand{\bc}{\begin{center}}
\newcommand{\ec}{\end{center}}

\newcommand{\bs}{${B}_s\rightarrow\mu^+\mu^-$}
\newcommand{\bd}{${B}_d\rightarrow\mu^+\mu^-$}
\newcommand{\bsd}{${B}_{s,d}\rightarrow\mu^+\mu^-$}
\newcommand{\cbs}{{\rm Br}({B}_s\rightarrow\mu^+\mu^-)}
\newcommand{\cbd}{{\rm Br}({B}_d\rightarrow\mu^+\mu^-)}
\newcommand{\dam}{\delta a_\mu}
\newcommand{\gsim}{\lower.7ex\hbox{$\;\stackrel{\textstyle>}{\sim}\;$}}
\newcommand{\lsim}{\lower.7ex\hbox{$\;\stackrel{\textstyle<}{\sim}\;$}}

\begin{document}
\tolerance=100000
  
\begin{flushright}
IPPP-04-41\\[-1mm]
DCPT-04-82\\[-1mm]
hep-ph/0407285\\[-1mm]
July 2004
\end{flushright}
  
\bigskip
  
\begin{center}

{\Large \bf Bounding the MSSM Higgs sector from above  
 }\\[0.3cm] {\Large \bf with the  Tevatron's  ${B}_{s} \to
 \mu^+ \mu^-$}\\[1.7cm]

{\large Athanasios Dedes$^1$ and B. Todd Huffman$^2$ }\\[0.7cm] {\it
$^1$Institute for Particle Physics Phenomenology (IPPP), Durham DH1
3LE, UK } \\[3mm] {\it $^2$University of Oxford, Particle Physics
Dept., Keble Road, Oxford OX1 3RH, UK}
\\[4mm]
\end{center}
 
\vspace*{1.8cm}\centerline{\bf ABSTRACT}
\vspace{0.1cm}\noindent{\small 

The discovery potential of the Tevatron CDF for the rare B-decay
${B}_{s} \to \mu^+ \mu^-$ is analysed. We find that with an integrated
luminosity of 2 fb$^{-1}$, and using CDF as the example detector, a
5$\sigma$ combined discovery reach of the Tevatron is possible if the
Branching ratio for ${B}_{s} \to \mu^+ \mu^-$ is $(1.7\pm 0.46) \times
10^{-7}$.  Such a possible signal for the decay ${B}_{s} \to \mu^+
\mu^-$ will invite large $\tan\beta$ values and set an upper bound on
the heaviest mass of the MSSM Higgs sector in a complete analogy to
the upper bound of the lightest observable supersymmetric particle
set from the excess over the SM prediction of the muon
anomalous magnetic moment.  If for example, the decay ${B}_{s} \to
\mu^+ \mu^-$ is found at Tevatron with $\cbs = 2\times 10^{-7}$ then
the heaviest Higgs boson mass in the MSSM should be less than 790~GeV
for $\tan\beta \lsim 50$ provided that the CKM matrix is the only
source for (s)quark flavour changing processes. }
 
\vspace*{\fill}
\newpage

\setcounter{equation}{0}
\section{Introduction}
\label{sec:intro}

In the Minimal Supersymmetric Standard Model (MSSM) the Higgs bosons
are {\it not} scalar superpartners of known Standard Model (SM)
fermions. Apart from a naturally light Higgs boson, the rest of the
Higgs scalars may well have masses at very high energies despite fine
tuning matters. We argue here that if the Tevatron observes \bs, 
the heaviest Higgs boson mass of the
MSSM is bounded from above to be less than 1 TeV for $\tan\beta \lsim
50$, provided that the Cabibbo-Kobayashi-Maskawa (CKM)
matrix~\cite{CKM} is the only source for (s)quark flavour changing
neutral current (FCNC) processes and CP-violation.

The published Tevatron/CDF physics results with  luminosity
$171~{\rm pb}^{-1}$ is the bound on \bs ~\cite{cdf2}
\be\label{exp}
\cbs < 5.8 \times 10^{-7} \;, \;\;\;\; {\rm at~~} 90\% ~{\rm C.L} \;.
\ee
The predicted ratio for $\cbs$ in the SM
is $(3.8\pm 1.0)\times 10^{-9}$~\cite{ILBB}.  The observable \bs~ makes
itself special in the MSSM where it is enhanced by orders of
magnitude. This enhancement is due completely to the MSSM neutral part
of the Higgs sector.  In particular, $\cbs$ is proportional to the
sixth power of $\tan\beta$ and inverse proportional to the fourth
power of the CP-odd (or CP-even) Higgs boson mass, $M_A$~\cite{MSSMbmumu}.
The idea here is that, since the parameter $\tan\beta$ is
theoretically\footnote{Large $\tan\beta \gsim 50-60$ values lead  to
non-perturbative evolution of the top or bottom Yukawa couplings at
higher energies.} bounded from above and the amplitude \bs~exhibits a
non-decoupling behaviour in the large SUSY mass limit, any observation
of \bs~at the Tevatron will set an upper bound on the neutral heavy Higgs
masses.  We present a quantitative analysis of this upper bound in the
MSSM under the assumptions of conserved R-parity symmetry and no extra
source for FCNC's other than the CKM matrix itself.

Although the rare decay \bs, and the SUSY contribution to the muon
anomalous magnetic moment $\dam$, go hand in hand in the minimal
Supergravity scenario~\cite{Dedes} this is not true in a general
unconstrained MSSM which we explore here. It is however possible, and
has been shown in Ref.\cite{Feng}, that following an argument along
the lines of the previous paragraph but for $\dam$, one could set an
upper mass bound on the lightest observable supersymmetric particle.
For complementarity purposes,
we repeat the analysis of Ref.\cite{Feng} with the up-to date 
(and persistently non-zero) measurement of $\dam$.

\setcounter{equation}{0}
\section{Discovery potential for $B_{s} \to \mu^+ \mu^-$ at Tevatron}
\label{sec:discovery}

In order for $B_{s,d} \to \mu^+ \mu^-$ to actually be observable at
the Tevatron it would need to appear as a clear, unambiguous signal
above background events.  Because this is an important signature,
potentially indicating unequivocally the existence of beyond the SM
physics, a requirement that any observed signal must be at least five
standard deviations above the level of background fluctuations would
likely be imposed by the experimenters.  Consequently, a study of the
discovery potential of any future experiment requires that we
estimate the amount of background in the experiment as the 
data increases. Because a detailed description of the background
levels, and their uncertainties, exists for the CDF collaboration,
what follows will draw heavily from their observed background
levels~\cite{cdf2,rade,fnalbrep}.

To estimate the number of signal events and precision obtained 
in a hypothetical future experiment, 
we assume that, with no changes in selection criteria for the events
from~\cite{cdf2}, 
the number of background events in the region of interest around the 
$B_{s,d}$ masses will be of the following form:
\begin{equation}
N_{bkg} = \sigma_{bkg} \cdot x \;,
\end{equation}
where $x = \int{\mathcal{L}\; dt}$ and $\sigma_{bkg}$ is the
equivalent cross section of the background in the mass region of
interest.  We set the
discovery level for the number of $B_{s,d} \to \mu^+ \mu^-$ candidates
($N_{sig}$) to be $N_{sig} \geq 5 \sqrt{N_{bkg}}$.
Using these relations we present the data in the form 
\begin{equation}
\sigma_{Bs,d} \cdot \rm{Br}(B_{s,d} \to \mu^+ \mu^-) = 
\frac{{N_{sig}}}{2 \epsilon \cdot \alpha \cdot x} \simeq
\frac{5\sqrt{\sigma_{bkg} \pm \xi(x)}}{2 \epsilon \cdot \alpha \sqrt{x}}\;,
\label{eq:expa}
\end{equation}
where $\epsilon \cdot \alpha$ is the 
efficiency of the entire selection process multiplied by the detector 
acceptance. The uncertainty in $\epsilon \cdot \alpha$
is independent of the integrated luminosity so this uncertainty 
has been included as part of the constant error on $\sigma_{bkg}$. 
$\xi(x)$ represents the fact that all experiments have
uncertainties and that often these uncertainties are based in part on
the data themselves. As a result we expect the precision on part of
the background estimate will improve as the data collects.
The factor of two accounts for the fact that the experiment does 
not distinguish between the charge conjugate decays.

The advantage of using equation~(\ref{eq:expa}) is that all the
quantities on the right are obtained from a single experimental
analysis.  In contrast, on the left side of the equation,
$\rm{Br}(B_{s,d} \to \mu^+ \mu^-)$ is the quantity of interest and
$\sigma_{Bs,d} = \frac{f_{s,d}}{f_u}\sigma(B^+)$ depends on $B$ meson
fragmentation measurements from other experiments and a cross section
measurement in a different decay mode (and for this paper, at a
different $\sqrt{s}$). So equation~(\ref{eq:expa}) allows the reader to
more easily adjust estimates of experimental reach for different
integrated luminosities, fragmentation estimates, and $B$ meson or $b$
quark cross section measurements.

Henceforth we will focus on the decay $\cbs$ only.  In $171$ pb$^{-1}$
the CDF experiment reports $N_{bkg} = 1.05 \pm 0.31$ events.  The 
largest
uncertainty is actually luminosity dependent because it
comes from the
statistical fluctuations of 14 events that remain in the unblinded
region after some of the final analysis cuts were relaxed.  $N_{bkg}$
can be converted into an equivalent background cross section using
$\sigma_{bkg} \pm \xi(x) = N_{bkg}/x_{0} = 6.14\times 10^{-3} \pm 2.9\times
10^{-4} \pm \frac{0.023}{\sqrt{x}}$ pb. For this paper $x_0 = 171$ pb$^{-1}$
while $\epsilon \cdot \alpha = 0.020\pm 0.002$.

Figure~\ref{fig3} shows how the $\sigma \cdot \cbs$ 
would scale with increasing integrated luminosity given that this decay
is sufficiently prolific to allow a discovery using parameters from
the CDF experiment. The uncertainty band shown here reflects the uncertainty
in the background estimate and takes into account the additional 
statistical uncertainty that would be added in quadrature if this decay 
was seen at the $5\sigma$ level.

Taking the $5\sigma$ values from Figure~\ref{fig3} for 0.5, 2.0, and 10
fb$^{-1}$ and using\footnote{ $f_x$ is the probability that a b-quark
fragments into a given $B$ meson.  The most recent results from the
Particle Data Group are used.  {\tt http://pdg.lbl.gov/}. }
$\frac{f_s}{f_u} = \frac{0.100}{0.391}$ and $\sigma(B+) = 3.6 \pm 0.6
\;\mu$b~\cite{cdfBplus} we obtain the results in Table~\ref{tab:a} for
the branching ratio $\cbs$.
\begin{table}[t]
  \centering
  \caption{ \it Tabulated are discovery branching ratios for CDF at three
different integrated luminosities.  The uncertainties are from the
background estimate in~\cite{cdf2}.}
  \vskip 0.1 in
  \begin{tabular}{|r|c|c|} \hline
    $\int{\mathcal{L}\: dt}$, fb$^{-1}$ & $\sigma_{Bs} \cdot 
\cbs$, pb 
				& $\cbs \times 10^{-7}$ \\
    \hline 
    \hline
    $ 0.5 $   & $0.44 \pm 0.14$ & $ 4.8 \pm 1.7 $  \\
     \hline
    $ 2.0 $   & $0.22\pm 0.05$ & $ 2.4 \pm 0.65 $ \\
    \hline
    $ 10.0 $   & $0.098 \pm 0.014$ & $ 1.1 \pm 0.24 $ \\
    \hline
  \end{tabular}
  \label{tab:a}
\end{table}
%
The analysis above makes use of the detailed reports on this decay
from the CDF experiment alone. In order to obtain a Tevatron estimate
one would need to include the D\O ~data as well. There is a
preliminary estimate for an upper limit of $\cbs = 4.6\times 10^{-7}$
at $95\%$ CL on D\O 's public analysis web site~\cite{d0:limit}.

It is difficult to account for all of the differences between the two
experiments, some of which improve and some of which detract from the
relative sensitivity between the two experiments, but it is clear that
a limit on this decay mode similar to CDF's has been
obtained. Consequently we felt justified in lowering our final
estimate of Tevatron reach by a factor of $\sqrt{2}$ but present only
the CDF-derived values in the paper because of a better understanding
of that detector by one of the authors.

\setcounter{equation}{0}
\section{Bounding the MSSM Higgs and SUSY sectors from above}
\label{sec:mssm}

As we stated in the introduction, excesses of the
experimental data over the SM expectation for observables like the
muon anomalous magnetic moment $\alpha_\mu$, and the rare B-meson decay \bs,
would result in bounding from above the  SUSY and the  Higgs sectors,
respectively. In this section,  we quantify these statements.

At large $\tan\beta$, the leading supersymmetric contribution to
$\cbs$ is depicted in Fig.(\ref{fig2}). Its amplitude depends mainly
on 6 beyond the SM parameters, namely, the CP-odd Higgs mass $M_A$ (or
the CP-even Higgs boson mass because $M_A\simeq M_H$), the ratio of
the vacuum expectation values, $\tan\beta$, the soft supersymmetry
breaking trilinear coupling $A_t$ and the two scalar superpartners of
the top quark masses, $m_{\tilde{t}_L}$ and $ m_{\tilde{t}_R}$.
It is well known~\cite{MSSMbmumu}, that the self energy diagram
of Fig.(\ref{fig2}), enhances the ratio $\cbs$ in the MSSM by many
orders of magnitude.  It is worth making a parenthesis here in order
to understand why this is so.
 
The leading contribution in $\cbs$,
comes from the amplitude  in  Fig.(\ref{fig2}). Its calculation 
 can be sketched with the following steps 
\begin{eqnarray}\label{appA1}
{\cal A}( b_R \to s_L \overline{\mu_R} \mu_L  )  
& \sim &  y_b \: \frac{1}{m_b} \:
y_b  \: V_{tb} \: \mu  \: y_t   \, A_t  \: v_u \: y_t \: V_{ts}^* \:
f(\mu,m_{\tilde{t}_L},m_{\tilde{t}_R}) \: \frac{1}{M_A^2} \: y_\mu
\\[0.4cm] 
\label{appA2} & \hookrightarrow & V_{tb} V_{ts}^* \: y_\mu \: y_b 
\: y_t^2 \:  \: \frac{\tan\beta}{M_A^2}
 \:  \biggl (\frac{\mu  \: A_t}{m_{\tilde{t}}^2} \biggr )
 \: \\[0.4cm] 
 \label{appA3}    & \hookrightarrow &
  V_{tb} V_{ts}^* \: m_\mu \: m_b \: m_t^2 \: \frac{\tan^3\beta}{M_A^2} 
\;,
\end{eqnarray}
where $V$ is the CKM matrix.  The function
$f(\mu,m_{\tilde{t}_L},m_{\tilde{t}_R})$ is a loop factor which, in the
limit of heavy squark masses, behaves like $1/m_{\tilde{t}}^2$. 
$m_{\tilde{t}}^2$ denotes  the geometric
mean squark mass, i.e $m_{\tilde{t}}^2 = m_{\tilde{t}_L}
m_{\tilde{t}_R}$.  This limit is taken when passing from Eq.(\ref{appA1}) to
Eq.(\ref{appA2}). The approximation symbol ``$\sim$'' stands for
constant, {${\cal O}(1)$, numerical factors irrelevant to our point
here.

Eq.(\ref{appA2}), shows that the amplitude is proportional to one
generic (i.e., not originated from the conversion of a Yukawa coupling
to a quark or lepton mass) $\tan\beta$ parameter.  The other two
powers of $\tan\beta$ arise in Eq.(\ref{appA3}) from the fact that the
Higgs field mediating the amplitude is the heavy one and at large
$\tan\beta$, is $[y_{b,\mu} \propto m_{b,\mu} \tan\beta]$.  The
reader should also notice that the light Higgs boson coupling to the
bottom quarks or to the leptons is {\it not} proportional to
$\tan\beta$ and thus leads to subdominant contributions.  It is very
interesting to note that, in the SUSY decoupling limit, $\mu \sim A_t
\sim m_{\tilde{t}} \to \infty$ in Eq.(\ref{appA2}), the ``Higgs
penguin'' {\it does not} decouple from the MSSM. This non-decoupling
property makes the rare decay \bs ~visible even for multi-TeV 
superparticle spectrum.

To summarise, squaring the amplitude in (\ref{appA3}) we obtain 
\begin{eqnarray}
\cbs \propto
~ \frac{\tan^6\beta}{M_A^4}\: 
\;.
\label{bmmapr}
\end{eqnarray}
Thus given an upper ``theoretical'' bound for $\tan\beta$ or measuring
$\tan\beta$ by other means,  an
observation of the decay \bs ~at Tevatron such as the one we described
in Table 1, will set an upper bound on $M_A$.

In our numerical analysis, we make use of two independent calculations
for $\cbs$.  The first contains the 1-loop calculation, including
corrections after the electroweak symmetry breaking~\cite{Bobeth} with
the method of $\tan\beta$ resummation as
described\footnote{Basically, one has to replace the two bottom
Yukawa couplings $y_b$, in Fig.(\ref{fig2}) with the effective
$y_b/(1+\Delta m_b)$ where $\Delta m_b \propto \mu \tan\beta$ depends
on the gluino and sbottom masses~\cite{Nierste}.  Notice that in
Eq.(\ref{appA2}) only one bottom yukawa coupling survives at the end.}
in~\cite{Dedes}.  Our second calculation is based on the
effective Lagrangian technique~\cite{Babu,Isidori,Buras,DP}. In particular
we follow the  work~\cite{DP} where a general resummed effective
Lagrangian for Higgs-mediated FCNC interactions in the MSSM has been
derived, without restrictions to particular assumptions that rely on
the quark-Yukawa structure of the theory or CP-conservation. For large
$\tan\beta$ the two calculations are in good agreement.

Our calculation of supersymmetric contributions to muon anomalous
magnetic moment $\dam$, follows the work in
Ref~\cite{Moroi}. Recently, two-loop $\tan\beta$-enhanced corrections
have been carried out~\cite{Dominik}. These corrections
account for $\dam \lsim 1 \times 10^{-10}$ if the sparticles are
heavier than approximately 400 GeV and they are neglected in our
analysis here.  Finally, for the Higgs mass bound we use the numerical
code from~\cite{Sven}.

We use a high statistics scan over the unconstrained
MSSM\footnote{Unconstrained MSSM means here that we impose no
restrictions on the boundary conditions for squark and slepton masses
or trilinear soft SUSY breaking couplings at the GUT scale. We do not
also require radiative electroweak symmetry breaking.} parameter space in the
region where the soft breaking masses are below 2.5 TeV.  We assume
that the FCNC couplings originate only from the CKM
matrix. Furthermore, we assume that the Lightest Supersymmetric
Particle (LSP) is stable and neutral\footnote{This is only necessary for
the R-parity conserved (we assume here) models.}.  That means
the LSP is either a neutralino or sneutrino. For each set of
outputs we define the Lightest Observable Supersymmetric Particle
(LOSP) as the second lightest SUSY particle or the third if the first
two LSP's are both neutralino and sneutrino. Finally, we scan only over
the positive sign of the $\mu$-parameter consistent with the bound
from $b\to s \gamma$. In any case, results for $\cbs \sim 10^{-7}$,
are independent of this assumption [see Eq.(\ref{appA2})]. The
unification assumption $M_1=M_2/2$ has been assumed for
simplicity. We have imposed the current experimental bounds from the
direct SUSY searches at LEP and Tevatron~\cite{booklet}. For the light
Higgs boson we have set a rather conservative constraint $M_h\ge 90$
GeV since we work in the large $\tan\beta$ regime.

Our results are summarised in Fig.(\ref{fig1}).  Fig.(\ref{fig1}a)
displays the ratio $\cbs$  with respect to the CP-odd Higgs mass, $M_A$.
Clearly, $\cbs$ can be three or even four orders of magnitude bigger
than the SM expectation and certainly within the 5$\sigma$ Tevatron/CDF
reach as shown in the figure, for $2~{\rm
fb}^{-1}$ (see Table 1 for other choices of luminosity).  Even for
very large $M_A\sim 1$ TeV, the ratio can still be enhanced by an order of
magnitude relative to its SM prediction.  However, as $M_A$ increases,
$\cbs$ can not be bigger than certain values. The envelope contour in
Fig.(\ref{fig1}a) is well approximated with
\begin{eqnarray}
\cbs &=& 5\times 10^{-7}~\biggl (\frac{\tan\beta}{50}\biggr )^6 
~\biggl (\frac{650~{\rm GeV}}{M_A^{}} \biggr )^4 
\ + \ 1.0\times 10^{-8} \;, \label{bmmnum1}
\end{eqnarray}
%
We have checked that Eq.(\ref{bmmnum1}) fits well for larger values of
$\tan\beta$.  Now, from Eq.(\ref{eq:expa}), if CDF with 2 ${\rm fb}^{-1}$
(10 ${\rm fb}^{-1}$) detects 17(88) signal events for the \bs that means
that $\cbs \approx 2\times 10^{-7}$ and the heaviest Higgs Boson mass will be
less than 790 GeV for {\it all} $\tan\beta \lsim 50$.

A more  ``natural''  prediction for $\cbs$
is approximated with the inner curve 
\begin{eqnarray}
\cbs &=& 5\times 10^{-7}~\biggl (\frac{\tan\beta}{50}\biggr )^6 
~\biggl (\frac{300~{\rm GeV}}{M_A^{}} \biggr )^4 
\ + \ 5.0\times 10^{-9} \;, \label{bmmnum2}
\end{eqnarray}
and represents the  statistically populated area of points.
Using  Fig.(\ref{fig1}a) and our analysis  
shown in Table 1 for  5$\sigma$ \bs 
~discovery at Tevatron/CDF, we conclude that for $\tan\beta \le 50$, 
\begin{eqnarray} \label{boundma}
M_A & \lsim & { 660}\: (305)~{\rm GeV}~~{\rm with}~~x=0.5~{\rm fb}^{-1} 
\; \nonumber \\[2mm]
M_A & \lsim & {790}\: (360)~{\rm GeV}~~{\rm with}~~x=2~{\rm fb}^{-1} \;
\nonumber \\[2mm]
M_A & \lsim & {970}\: (440)~{\rm GeV}~~{\rm with}~~x =10~{\rm fb}^{-1}
\;, \; 
\end{eqnarray}
where the numbers in parenthesis fall into the Eq.(\ref{bmmnum2}).
Further uncertainties 
due to the $B_s$ decay constant $f_{B_s}$, are also reduced
when calculating the $M_A$, as compared to the ones involved in 
the calculation of $\cbs$, since it is $\sqrt{f_{B_s}}$ vs. $f_{B_s}^2$.
An error of 5-10\% on $M_A$ can still be present but can be
further reduced once $\Delta M_s$ is measured, possibly at Tevatron,  
along the lines of the last paper in Ref.\cite{ILBB}.
Eq.(\ref{boundma}) is the main result of our  paper.

The current BNL~\cite{BNL} experimental result on the muon anomalous
magnetic moment, shows a 2.7$\sigma$ difference relative to the SM
expectation. The deviation from the SM value $\dam =\alpha_\mu^{\rm
exp}-\alpha_\mu^{\rm SM}$ at 90\% C.L reads~\cite{Nomura}
\be
12.9\times 10^{-10} \lsim \dam \lsim 36.0 \times 10^{-10} \;,
\label{ambound}
\ee
with a central value $\dam = 24.5 \times 10^{-10}$~\cite{Nomura}~\footnote{
New estimates on the light by light scattering~\cite{estimates} as
well as QED $\alpha^4$ contributions~\cite{QED} have been taken into
account when quoting this value.}.  A
possible excess can originate from six  supersymmetric mass parameters,
namely, the gaugino masses $M_1$ and $M_2$, the Higgsino mass
parameter $\mu$, two supersymmetric scalar partner masses of the muons,
$m_{\tilde{\mu}_L}$ and $m_{\tilde{\mu}_R}$,  and $\tan\beta$. Then, in
the mass insertion approximation, one may write this contribution
simply as~\cite{Feng}
\begin{eqnarray}
\dam \ \simeq \ \frac{g_i^2}{16\pi^2}\: m_\mu^2 \: \mu \: M_i \:
 \tan\beta \: F \;,
\label{g-2}
\end{eqnarray}
with $F\propto 1/M_{SUSY}^4$ being a loop factor depending on the
supersymmertic  masses (charginos, neutralinos, smuons).   
Feng and Matchev in~\cite{Feng}  used the
excess on the muon anomalous magnetic moment to set an upper
mass bound on the Lightest Observable Supersymmetric Particle ($M_{\rm LOSP}$).
 We repeat their analysis in Fig.(\ref{fig1}b), where we plot the
 supersymmetric contribution $\dam$, versus the mass of the LOSP.
 The current excess on $\dam$ already sets an upper bound
$M_{LOSP}\lsim 700$ GeV, on the LOSP mass. This is clear from the
 envelope solid contour in Fig.(\ref{fig1}b), which is described by
 the equation
\be
\dam = 18\times 10^{-10} ~ \frac{\tan\beta}{50} 
~ \biggl (\frac{550~{\rm GeV}}{M_{\rm LOSP}^{}} \biggr)^2 \;.
\ee
Thus the current central value~\cite{BNL,Nomura}
$\dam = 24.5\times 10^{-10}$  implies that 
the lightest  observable
Supersymmetric particle should weigh less than 450 GeV. 
The analogy  with the ratio $\cbs$ in Fig.(\ref{fig1}a) is obvious.

Notice that, from the Fig.(\ref{fig1}a) one cannot set a lower bound
on the CP-odd Higgs mass from the Tevatron/CDF experimental bound
alone. The parameter which can be excluded by CDF is the ratio of the
soft SUSY breaking trilinear parameter with the CP-odd Higgs mass,
$A_t/M_A$.  This bound can be understood easily from Eq.(\ref{appA2})
and can be derived from Fig.(\ref{fig1}c) to be
\be
\frac{|A_t|}{M_A} \lsim 12 ~~{\rm for}~~\tan\beta\gsim 50\;.\;
\label{atbound}\ee
Combining Eqs (\ref{atbound}) and (\ref{boundma}) we arrive at the
conclusion that any observation of $\cbs$ will not only bound the
Higgs sector from above but will also set an upper bound to the top
trilinear soft SUSY breaking coupling.  A soft supersymmetry breaking
parameter is bounded by a Higgs sector parameter!
 
Finally, in Fig.(\ref{fig1}d) we plot the two observables $\dam$ and
$\cbs$ together.  Since both are saturated by different mass scales
[$\dam$ by the SUSY and $\cbs$ by the Higgs mass scale)], we do not
expect any correlation between them in the general unconstrained MSSM.
This is exactly what we obtain in Fig.(\ref{fig1}d). In fact, the
statistical distribution of the points with large $\cbs$ and small
$\dam$ is wider than the other way around (we have scanned the
parameter space uniformly).  However, in the mSUGRA scenario, the SUSY
mass $M_{LOSP}$ and the CP-odd Higgs mass, $M_A$, are related
(mainly) through the parameter $M_{1/2}$. The mSUGRA points in
Fig.(\ref{fig1}d) lay along the diagonal in the $\dam-\cbs$ plane.  As
it was remarked in~Ref.\cite{Dedes}, a discovery of $\cbs$ at
the Tevatron and a deviation on $a_\mu$ (as is currently the 
case~\cite{SUSYtalks}) on both observables with respect to their SM
values will definitely favour some kind of mSUGRA scenario.

We have checked the robustness of our results presented in
Fig.(\ref{fig1}) against {\it i)} a general scan in the multi-TeV
region of soft supersymmetry breaking masses, {\it ii)} inclusion of
various supersymmetric phases and {\it iii)} the inclusion from the gluino
loop corrections. The last two contributions have been implemented in
the analysis of Ref.\cite{DP}. We find numerically that these
contributions do not spoil the upper bound of Eq.(\ref{boundma}). 
However one expects different bounds in the case of non-trivial squark
mass intergenerational mixings in the supersymmetry breaking
sector. The effects of the latter will not be missed at B-factories,
colliders, or other non-accelerator experiments; where strong
constraints on them already exist.

We note in passing, that Tevatron~\cite{cdf2} and
B-factories~\cite{Bfactories} search for the $B_d$ leptonic decay,
\bd.  Current limits obtained in the region $[{\rm few} \times
10^{-7}]$. The branching ratio $\cbd$ is smaller than $\cbs$ by a
factor $|V_{td}/V_{ts}|^2 \lsim 0.06$.  Any observation at Tevatron
for \bs~is likely to be accompanied with an observation of \bd~at
B-factories if the Higgs sector is responsible for making these leptonic
rare decays visible.

\setcounter{equation}{0} 
\section{Epilogue}
\label{sec:sum}

We investigate the discovery potential of the Tevatron hadron
collider, focusing mainly on the CDF experiment, for the rare decay
\bs. We find that Tevatron can reach a 5$\sigma$ (3$\sigma$)
observation for $B_{s}$-cross section times $\cbs$ with luminosity
given in Fig.(\ref{fig3}). Based on these results and the knowledge of
the $\sigma(B_s)$ from the Tevatron we show in Table 1 the 5$\sigma$
discovery bands for $\cbs$ at several luminosities. For example, the
Tevatron/CDF with integrated luminosity 2 ${\rm fb}^{-1}$, can
discover \bs ~with a ratio $\cbs = (2.4 \pm 0.65) \times 10^{-7}$. If
this is true, we show that in the MSSM, with the assumption that
flavour changing processes originate exclusively from the CKM matrix,
the Higgs sector is bounded from above with the heavy Higgs bosons
weight no more than 790 GeV for $\tan\beta \lsim 50$ [see
Fig.(\ref{fig1}a) and Eq.(\ref{boundma})]. The situation is completely
analogous with the muon anomalous magnetic moment searches, where the
current excess over the SM prediction seems to indicate an upper bound
of 700 GeV on the lightest observable supersymmetric particle.

\subsection*{Acknowledgements}
AD would like to thank P. Zerwas for illuminating discussions on the
matter, and useful communication with D. Nomura on Eq.(\ref{ambound}).
AD would also like to thank D.~St\"ockinger for discussions
on~Ref.\cite{Dominik}.  BTH would like to thank D. A. Glenzinski for
his help with many of the details of the CDF detector response to \bs.


\newpage

\begin{figure}[t]
\centerline{\hbox{\psfig{figure=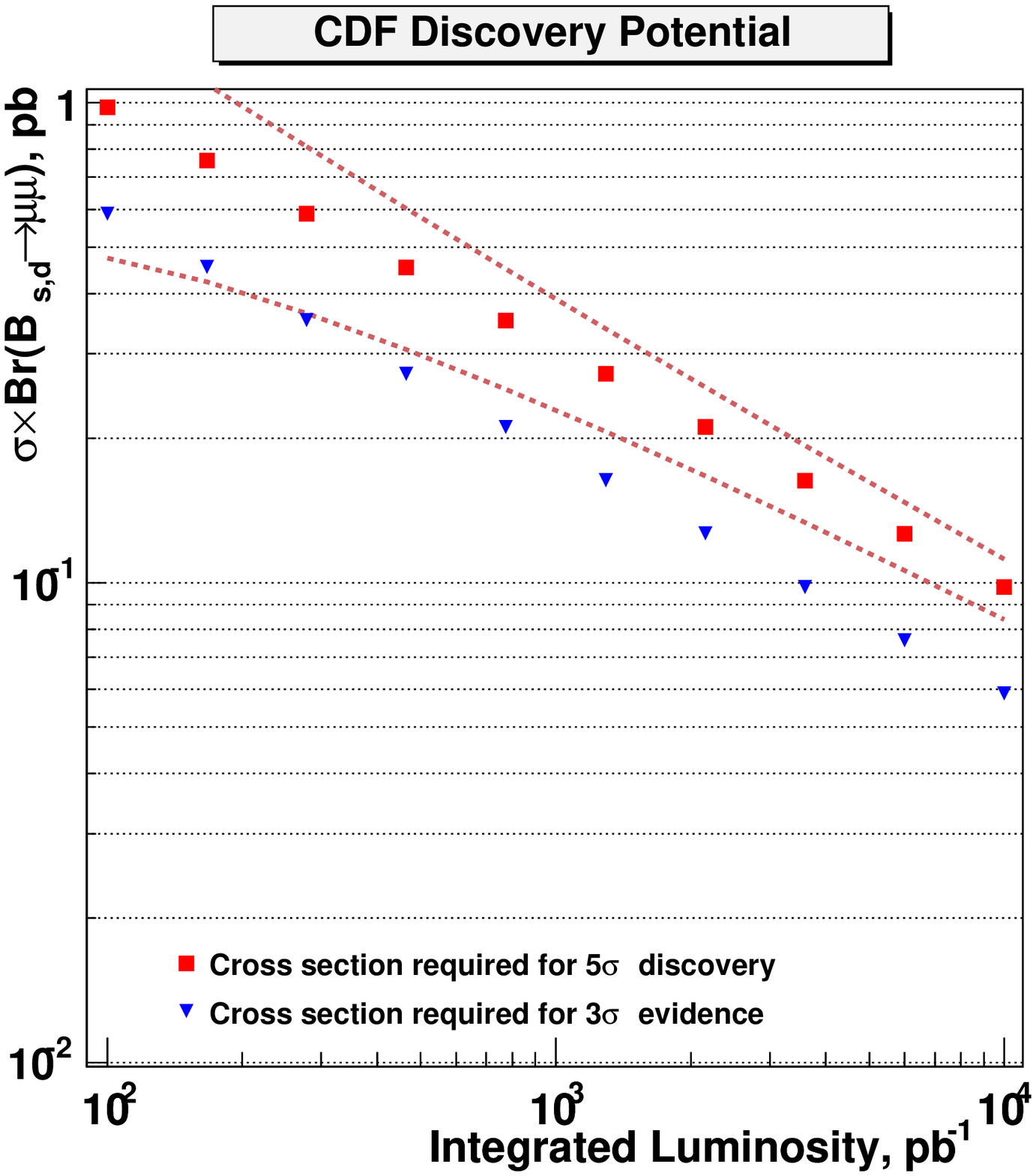,height=6in}}}
\caption{{\it Tevatron's CDF discovery potential for the product of
the production cross section times the Branching ratio
$[\sigma(p\bar{p} \to B_{s,d}+X) * \cbs]$ vs. integrated Luminosity.
The red (blue) squares are the center points for $5\sigma~(3\sigma)$
observation. The dashed lines delineate the uncertainty in 
the discovery cross section (both statistical and systematic)
as a function of integrated 
luminosity.}}\label{fig3}
\end{figure}

\begin{figure*}[t]
\centerline{\psfig{figure=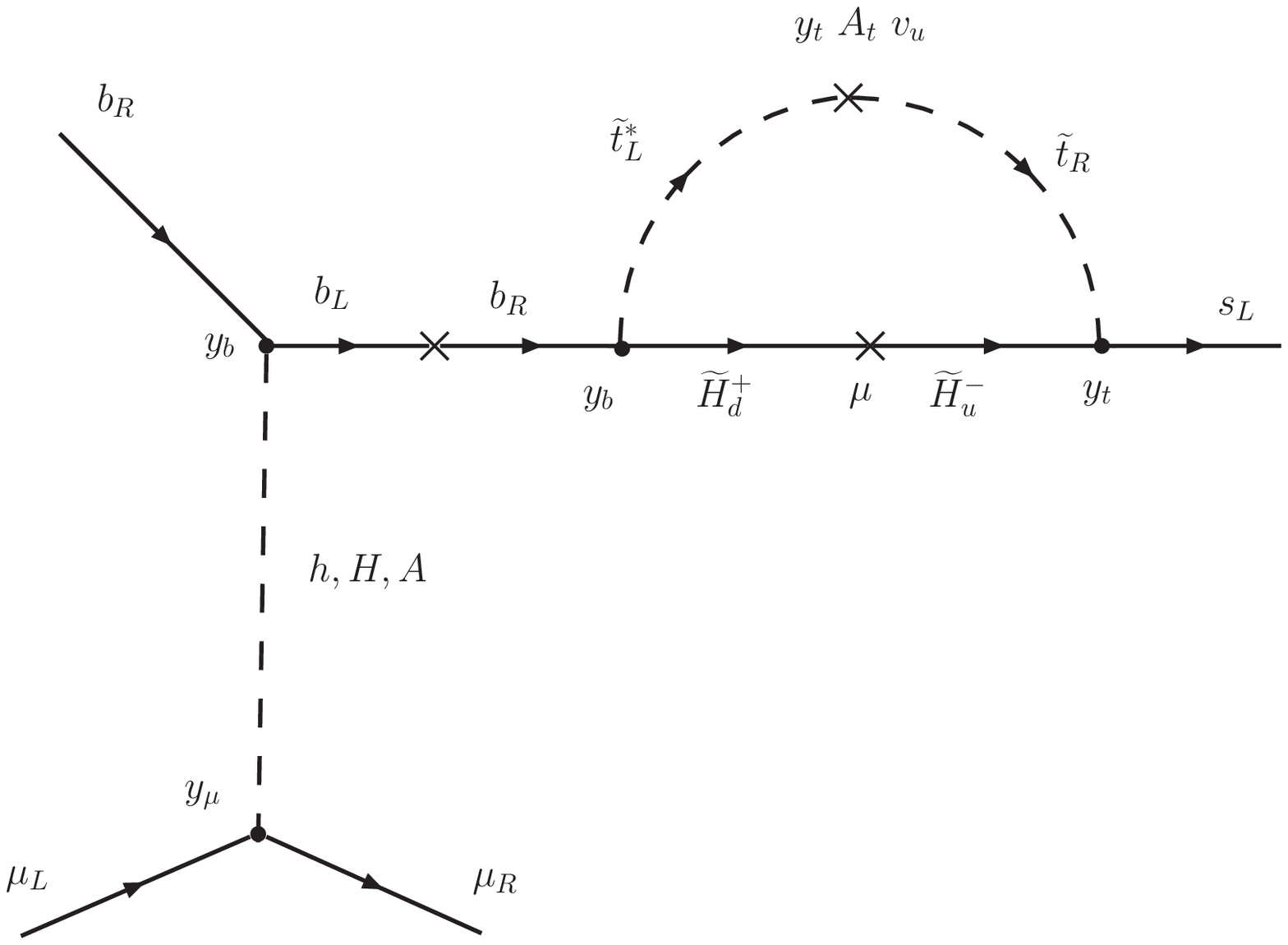,width=6in}}
\caption{{\it The leading supersymmetric contribution to \bs at large
$\tan\beta$.  ``Crosses'' and ``dots'' indicate mass insertions 
and  Yukawa vertices, respectively, and their  
corresponding interaction parameters are explicitly given.}  }
\label{fig2}
\end{figure*}

\begin{figure*}[ht]
\centerline{\psfig{figure=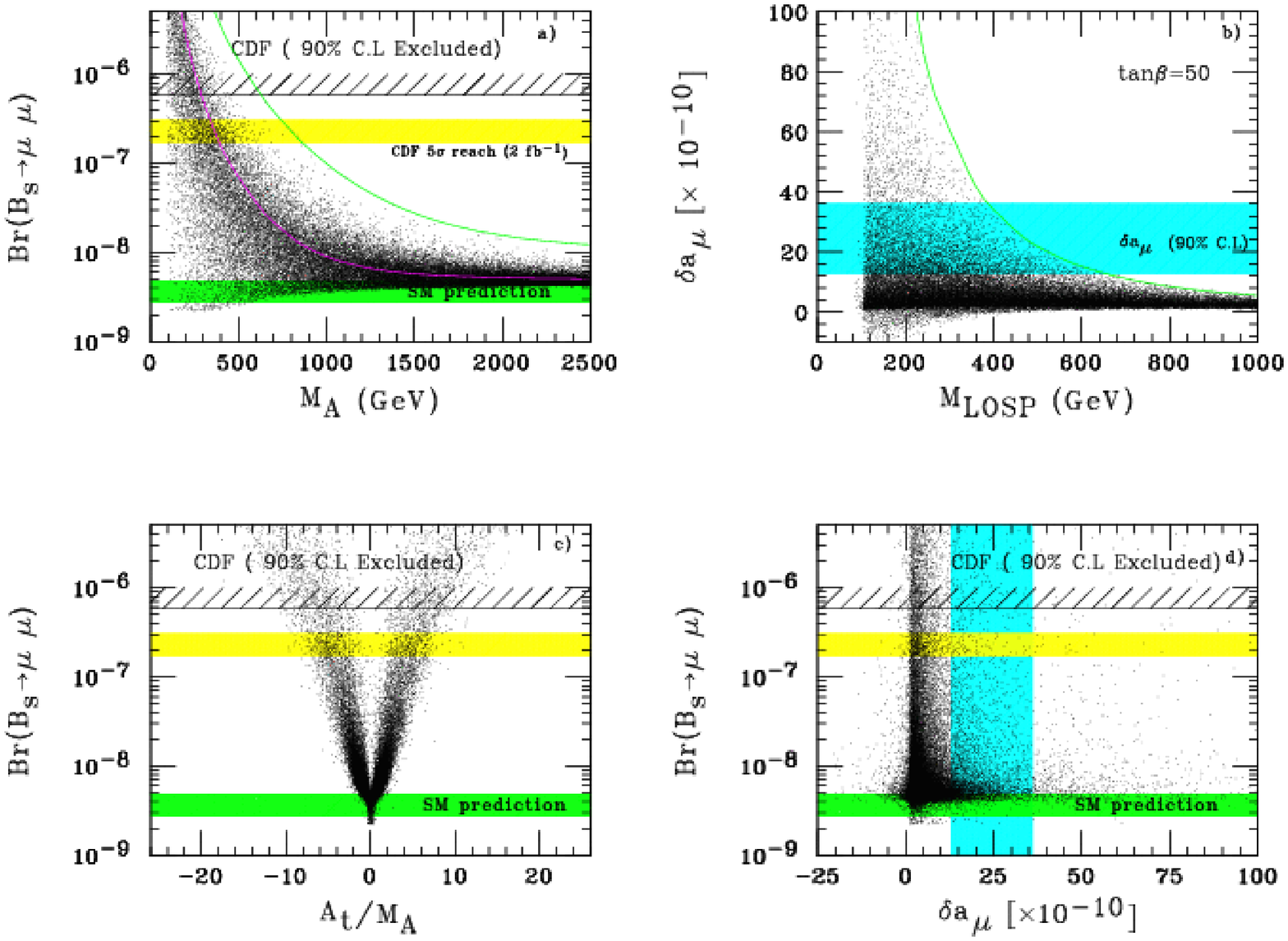, angle=90,width=7in}}
\caption{ {\it {\bf a)} The $\cbs$ vs. the CP-odd Higgs mass in the
MSSM. The lower shaded  band shows the SM prediction for $\cbs$.  The
90\% C.L excluded area from Tevatron/CDF~\cite{cdf2} is also
displayed.  Following our results from Table 1, the 5$\sigma$ reach at
2$~{\rm fb}^{-1}$ for $\cbs$ is also depicted (upper shaded band). 
 {\bf b)} The SUSY
contribution to the muon anomalous magnetic moment, $\dam$ versus the
mass of the Lightest Observed Supersymmetric Particle
(LOSP). Black points correspond to neutralino or sneutrino LSP. The
shaded region is the allowed value on $\dam$ at 90\%
C.L~\cite{BNL,Nomura}.  {\bf c)} The $\cbs$ vs. the ratio of the soft
SUSY breaking trilinear parameter with the CP-odd Higgs mass,
$A_t/M_A$.  {\bf d)} Combined predictions for $\dam$ and $\cbs$ in the
MSSM. For all the figures here we have chosen $\tan\beta=50$ and
$m_t=178$ GeV. }}
\label{fig1}
\end{figure*}

\end{document}